\documentclass[journal]{IEEEtran}
\usepackage{cite}
\usepackage[pdftex]{graphicx}
\usepackage[pdftex]{color}
\graphicspath{/images/}
\usepackage[ruled,vlined,linesnumbered]{algorithm2e}
\usepackage[cmex10]{amsmath}
\usepackage{algpseudocode}
\usepackage{array}
\usepackage{url}
\usepackage{mathtools}
\usepackage{color}
\usepackage{amsmath, amssymb}
\usepackage{amsthm}
\usepackage{booktabs}
\usepackage{balance}

\begin{document}
\title{Reliable and Efficient Access for \\ Alarm-initiated and Regular M2M Traffic in \\ IEEE 802.11ah Systems}

\author{Germ\'an Corrales Madue\~no, \v Cedomir Stefanovi\' c, Petar Popovski
\thanks{The authors are with the Department of Electronic Systems, Aalborg University, Aalborg, Denmark (e-mail: \{gco,cs,petarp\}@es.aau.dk).}
\thanks{The research presented in this paper was supported by the Danish Council for Independent Research (Det Frie Forskningsr\aa d), grant no. 11-105159 ``Dependable Wireless~bits for Machine-to-Machine (M2M) Communications'', grant no. 4005-00281 ``Evolving wireless cellular systems for smart grid communications'' and by the European
Research Council (ERC Consolidator Grant Nr. 648382 WILLOW) within the
Horizon 2020 Program.}
\thanks{Copyright (c) 2012 IEEE. Personal use of this material is permitted. However, permission to use this material for any other purposes must be obtained from the IEEE by sending a request to pubs-permissions@ieee.org.}
}

\maketitle

\begin{abstract}
	
	IEEE 802.11ah is a novel WiFi-based protocol, aiming to provide an access solution for the machine-to-machine (M2M) communications. 
	In this paper, we propose an adaptive access mechanism that can be seamlessly incorporated into IEEE 802.11ah protocol operation and that supports all potential M2M reporting regimes, which are periodic, on-demand and alarm reporting.
	The proposed access method is based a periodically reoccurring pool of time slots, whose size is proactively determined on the basis of the reporting activity in the cell.
	We show that it is possible to both efficiently and reliably resolve all reporting stations in the cell, within the limits of the allowed deadlines.
	As a side result, we also provide a rationale for modeling the inter-arrival time in alarm events by using the Beta distribution, a model that is considered in the 3GPP standardization.

\end{abstract}

\IEEEpeerreviewmaketitle

\section{Introduction}
\label{sec:intro}
	
	IEEE 802.11ah is a novel wireless local area network protocol, offering low-power, long-range and scalable operation.
	The operating  frequency of IEEE 802.11ah is below 1~GHz, allowing a single access point (AP) to provide service to a area of up to 1~km.
	Additionally, IEEE 802.11ah is designed to support very large number of stations (up to 8100) in a single cell, in comparison to the other standards from the IEEE 802.11 family.
	As such, IEEE 802.11ah aims to establish itself as an access solution with a low infrastructure cost for capillary M2M access.  
	
	At the time of writing, the IEEE 802.11ah task group is still working on the completion of the protocol definition, where the second draft was initially expected by mid-2014 \cite{qualcommProjects}.
	Nevertheless, the foreseen EEE 802.11ah access mechanism has already been topic of several recent studies.
	The number of devices supported for different uplink-downlink traffic ratios with different data rates was investigated in \cite{capacityAnalysis}.
	The performance of the grouping strategy, labeled as restricted access window (RAW), was studied in \cite{groupStrategy}.
	It was shown that limiting the number of contending stations through use of RAW could produce significant throughput improvements.
	In \cite{estimationRAW}, the authors propose to estimate the number of contending devices in order to determine the adequate length of the RAW.
	
	The main figure of interest in the above studies is the throughput, where the analysis is based on a full-buffer approach, i.e., it is assumed that the stations always have data to transmit.
	However, this assumption does not correspond to the nature of typical reporting from M2M devices (terminals), where the terminals only sporadically report data, triggered by the reporting period or certain external events.
		
	In contrast to previous works, in this paper we consider the reporting regimes that are characteristic for M2M, which differ significantly from the approach that assumes full buffer. 
	The first regime is the \emph{asynchronous periodic} reporting, characterized with reporting intervals of several minutes \cite{Madueno2017}. 
	The second regime is \emph{synchronous alarm} reporting, where potentially thousands of stations are triggered almost simultaneously.
	The showcase example of an M2M application in which terminals report in both regimes is smart metering.
	Under normal conditions, smart meters asynchronously and periodically report the energy consumption to a remote server.
	However, in special occasions, such as in a power outage, thousands of devices almost simultaneously try to report the failure before the battery dies \cite{lastGasp}.
	
	The topic of this paper is the design of an access method tailored for IEEE 802.11ah protocol, whose main aim is the provision a reliable and efficient service, both for asynchronous and synchronous M2M reporting.
	The proposed solution consists of the periodically reoccurring pool of resources, i.e., time slots, which is proactively dimensioned based on the outcomes of the contention among the reporting terminals and operates such that the dominant reporting regime, i.e., asynchronous vs. synchronous, is matched.
	Moreover, in contrast to previous works, we consider both the efficiency and reliability of the access mechanism, as the latter performance parameter plays an important role in M2M communications, particularly in alarm reporting scenarios.
	We show that it is possible to detect all alarms reports while efficiently using the system resources, i.e., wireless link time.
	The proposed scheme draws inspiration from the work presented in \cite{WCL}, where the main topic is a static solution for provision of a reliable communication to a set of M2M terminals. However, the scenario is limited only to the case in which the set of M2M terminals report asynchronously. Differently from \cite{WCL}, here we thoroughly modify and extend the solution from \cite{WCL} towards a dynamic version that is adapted to the reporting demands in a cell. Specifically, the proposal in this paper provides an efficient reliable support of synchronously reported alarms, which is a very challenging traffic class.
	
	Another important contribution of our work is related to the Beta distribution, used to model alarms that are reported synchronously. The use of the Beta distribution has been considered in standardization \cite{TR37.8682011}, however, without rationale on the choice of the distribution and its parameters. This paper closes this knowledge gap and provides a systematic treatment of the Beta distribution in relation to the M2M traffic.
		
	The rest of the paper is organized as follows.
	Section~\ref{sec:protocol} provides an overview of the 802.11ah access mechanism.
	Section~\ref{sec:model} presents the system model.
	Section~\ref{sec:trafficModel} is devoted the modeling of alarm reporting.
	Section~\ref{sub:proposedAllocation} introduces the proposed access method.
	Section~\ref{sec:analysis} is the central part of the paper, presenting analysis of the proposed scheme.
	The evaluation of the scheme's performance is done in Section~\ref{sec:results}.
	Finally, Section~\ref{sec:conclusion} concludes the paper.

\section{Basics of IEEE 802.11ah Access}
\label{sec:protocol}

\begin{figure}
	  \centering
	    \includegraphics[width=\columnwidth]{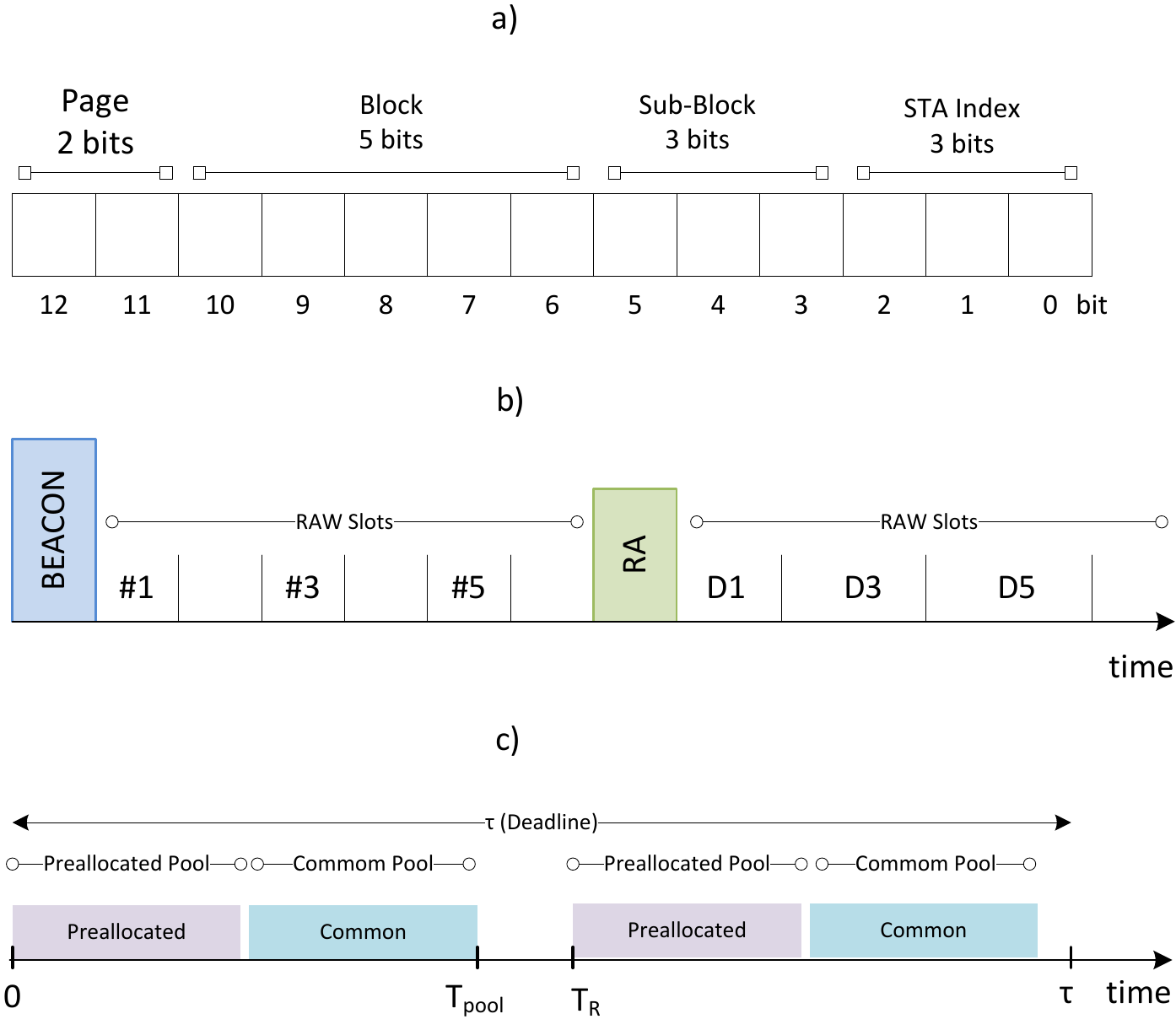}\caption{a) AID format. b) Beacon frame announcing a RAW with 6 slots followed by a Resource Allocation (RA) frame that allocates data slots for stations 1, 3 and 5. c) Illustration of $T_\text{R}$, $T_\text{pool}$ and the deadline $\tau$.}
	    \label{AID}
	\end{figure}

	The IEEE 802.11ah access mechanism operates as follows.
	Every station (i.e., terminal) connected to the AP receives an unique identifier, denoted as an association identifier (AID).
	The AID is a 13-bit word that follows a four-level hierarchical structure as depicted in Fig.~\ref{AID}a).
	The first two bits are used to organize the stations in four pages.
	The next 5 bits split a page into 32 blocks.
	The following three bits are used to divide a block into 8 sub-blocks.
	Finally, the last 3~bits are used to determine the station's index within the sub-block, i.e., there could be a maximum of 8 stations per sub-block.
	Obviously, this hierarchical structure allows for a straightforward grouping of stations.
	
	The AID can be used to determine which stations are allowed to access the medium.
	Specifically, IEEE 802.11ah introduces the concept of restricted access window (RAW), during which only certain stations are allowed to contend based on their AIDs.
	This limitation the number of contending stations in RAW enhances the access efficiency, seen in the number of stations whose transmissions do not experience collisions.
	RAW structure is depicted in Fig.~\ref{AID}b), where it can be seen that RAW is subdivided in slots.
	The information concerning the number of slots in the RAW, slot duration and assignment is indicated in the beacon frame that is periodically sent by the AP to the stations.
	Every slot can be assigned to a single or multiple stations, according to the number of stations and number of the available slots \cite{uplinkAccess}.
	The active stations contend only in the assigned slots.
	Henceforth, we assume that the contention is based on slotted ALOHA \cite{R1975}, where the contention outcomes can be an idle, singleton or collision slot, and that the cross-boundary slot option \cite{qualcommProjects} is not used, implying that a station is allowed to contend only within the boundaries of the allocated slot.
	The amount of information that can be sent within a slot depends on the slot duration and the modulation used; the available modulation schemes range from BPSK to 256-QAM and the corresponding data rates from 150~kbps to  4000~kbps \cite{qualcommProjects}.

	Two main RAW operation modes are generic RAW and triggered RAW.
	In case of the generic RAW mode, the stations contend with their data packets.
	In the triggered RAW mode, the access is reservation based and there are actually two RAWs, the initial RAW corresponding to the reservation phase and the second RAW corresponding to the data phase.
	In the initial RAW, the stations contend with short poll frames\footnote{At the time of writing, the structure of the poll frame has not been fully standardized.}, trying to reserve (i.e., obtain from the AP) a dedicated RAW slot for the pending downlink/uplink data transmissions.
	Based on the received polls, the AP allocates a second RAW consisting of dedicated slots, and notifies the corresponding stations via resource allocation frame (RA) about the RAW slot assignment, see Fig.~\ref{AID}b).
	The advantage of triggered over the generic RAW mode is that idle or collided slots are less costly due to their shorter length, as they just have to fit a poll frame instead of a data packet.

	For a better understanding, we provide an example of the triggered RAW mode in Fig.~\ref{AID}b).
	The AP provides an initial RAW composed of 6 RAW slots intended for 6 stations, i.e., a slot per station.
	It is assumed that only stations \#1, \#3 and \#5 have data to transmit and thus send their polls, whereas \#2, \#4 and \#6 are idle and do not transmit.
	Immediately after the initial RAW, the AP allocates another RAW containing a slot dedicated to each station, denoted as D1, D3 and D5 in Fig.~\ref{AID}b), and informs the stations about the allocation via RA frame.
	
	We conclude this section by noting that proposed access method is tailored for the triggered RAW mode of operation.
	In particular, the proposed method is intended as the efficient and reliable solution for the reservation phase of the triggered RAW mode.
	
\section{System Model}
\label{sec:model}

	We consider a single IEEE 802.11ah cell. 
	We assume that the cell is circular, with radius $r$, and that the AP is residing in its center.
	There are $N$ stations in the cell, whose distances from the AP are uniformly randomly distributed.
	We consider the scenario where the locations of the stations are fixed. This is a widely used assumption for M2M deployment based on scenarios with smart meters and sensors; this is also why one of the main changes to the cellular standards \cite{Madueno2017}, in order to make them suitable for M2M communications, is to remove the mobility-related features in the protocol.

	All stations perform periodic, on-demand and alarm reporting.
	\emph{Periodic reporting} refers to the automatic reporting performed by stations, which is modeled by a Poisson arrival process on a station basis with rate $\lambda_\texttt{p}=1/T_\text{RI}$, where $T_\text{RI}$ denotes the duration of the reporting interval (RI) \cite{WCL}.
	In the smart-metering context, periodic reporting is the automatic consumption reporting performed by the meters.
	\emph{On-demand reporting} is the reporting that is demanded by the application server and can be also modeled as a Poisson arrival process with rate $\lambda_\texttt{d}$; it could be expected that $\lambda_\texttt{p} > \lambda_\texttt{d}$ \cite{openSmartGrid}.
	An example for on-demand reporting is a delivery of consumption reports triggered by the customer.
	Both the periodic and on-demand reporting are asynchronous; henceforth refer to their combination as \emph{regular} reporting, as it corresponds to a regular, i.e., standard reporting in the context of M2M applications, when there are no alarm events.
 	Finally, \emph{alarm reporting} corresponds to traffic generated by an event in which affected devices are activated almost simultaneously.
	Modeling of alarm reporting is significantly more involved than in the previous two cases; we devote the following section to the derivation of the appropriate model.
	
	A parameter of paramount importance is the maximum allowed delay, i.e., the deadline from the report generation until its delivery.
	A report whose delay exceeds the maximum allowed value is considered as outdated and dropped by the station.
	Typically, it is assumed that regular reporting is more delay tolerant, i.e., a report in the current RI could be delivered in the beginning of the following RI \cite{Madueno2017}.
	Therefore, the maximum allowed delay of the periodic reporting is:
	\begin{align}
	\label{eq:periodic}
	\tau_\texttt{p} = T_\text{RI}.
	\end{align}
	On the other hand, on-demand and, particularly, alarm reporting have stricter deadlines \cite{openSmartGrid}:
	\begin{equation}
	\label{eq:deadlines}
		\tau_\texttt{a} \ll \tau_\texttt{d} \leq \tau_\texttt{p},
	\end{equation}
	where $\tau_\texttt{a}$ and $\tau_\texttt{d}$ denote alarm and on-demand reporting deadlines, respectively.
	
\section{Reporting of Alarms}
\label{sec:trafficModel}
			
	The typical assumption is that in case of an alarm event a large number of stations will be affected and their reporting will become correlated in time.
	In \cite{TR37.8682011}, 3GPP proposed a model for highly correlated traffic arrivals, where the interval-arrival time follows a Beta distribution:
	\begin{equation}
	\label{eq:beta}
		p(t) = \frac{t^{\alpha - 1} (T-t)^{\beta -1}}{T^{\alpha+\beta-1} \text{B}(\alpha,\beta)} \text{ for 0 $\leq t \leq T$},
	\end{equation}
	where $\alpha > 0$ and $\beta > 0$ are shape parameters, $\text{B}(\alpha,\beta)$ is the Beta function \cite{betaFunction} and $T$ is the activation period, i.e., the time period from the activation of the first station until the last station is activated.
	The number of stations triggered during an interval $T_I$ is given by:
	\begin{equation}
		N_A = N \int_{0}^{T_I} p(t) dt,
	\end{equation} 
	where $N$ is the total number of stations.
	
	The values of the parameters $\alpha$, $\beta$ and $T$ suggested in \cite{TR37.8682011} are $\alpha = 3$, $\beta = 4$ and $T=10 \, \text{s}$. However, reference \cite{TR37.8682011} neither elaborates on the reasons to use Beta distribution, nor provides a rationale for the choice of the values of the parameters. Our present work closes 
this knowledge gap as follows: modeling the reporting behavior of the stations, providing a justification for the use of Beta distribution for the characterization of the alarm reporting regime, providing insights into the choice of the parameters $\alpha$, $\beta$ and $T$, as well as their relation to the values suggested by 3GPP.

	\subsection{The Proposed Model of Alarm Reporting}
	\label{sec:reporting model}

\begin{figure}
			\centering
		   	\includegraphics[width=0.8\columnwidth]{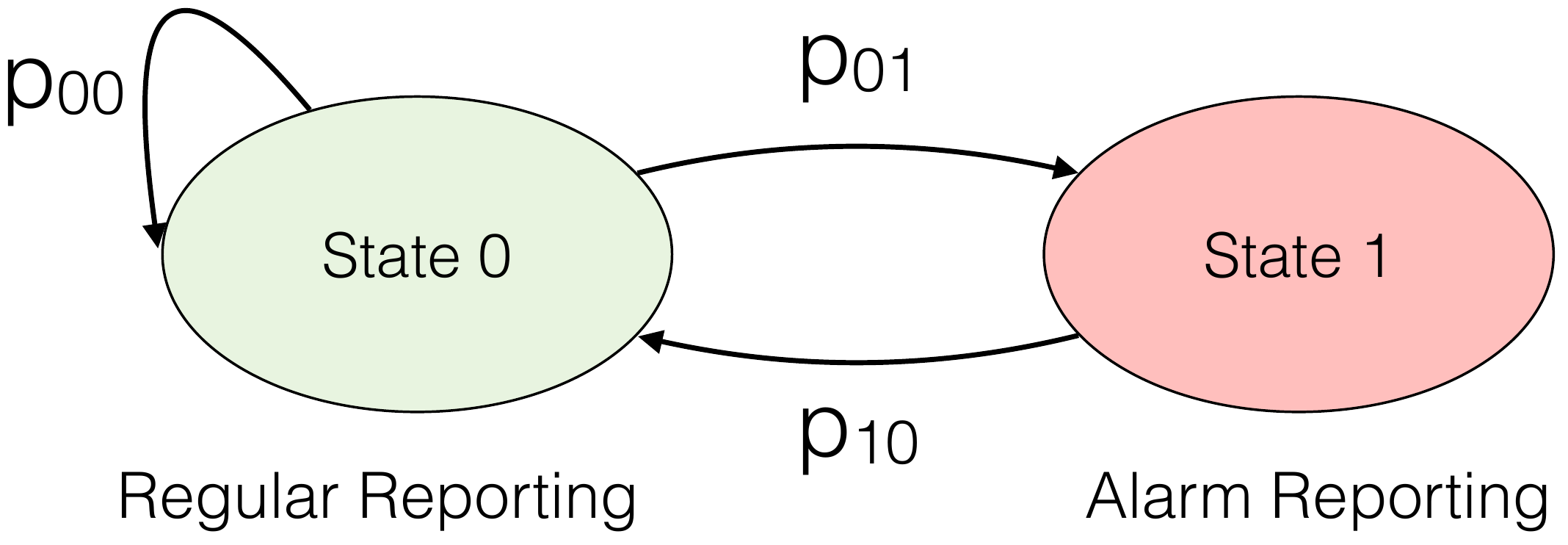}
	   		\caption{State diagram of a station, with two states that correspond to regular reporting state (state 0), comprising periodic and on-demand reporting, and alarm reporting state (state 1).}
	   	\label{fig:states}
	   	\end{figure}

		We assume that every station can be in two reporting states, see Fig.~\ref{fig:states}.
		The first, regular reporting state, denoted as state $0$, comprises periodic and on-demand reporting, and the second state, denoted as state $1$, corresponds to alarm reporting.
		 We assume that the state transitions occur in discrete time, where time step is denoted as $\Delta t$.\footnote{To appropriately  model the alarm reporting with respect to the proposed access method, it is sufficient that $ \Delta t \leq T_\text{R}$, where $T_\text{R}$ is the pool reoccurring period, see Section~\ref{sub:proposedAllocation}.}
		The arrival rates for each state during $\Delta t$ are:
		\begin{align}
		\lambda_0  = ( \lambda_\texttt{p}   + \lambda_\texttt{d}) \Delta t, \label{lambda0}
		\end{align}
		where $\lambda_\texttt{p}$  and $\lambda_\texttt{d}$ are arrival rates of the periodic and on-demand reporting, which are adjusted to the time step $\Delta t$ in \eqref{lambda0}.
		For analytical tractability, we also assume Poisson arrivals in state $1$  \cite{trafficModels}, with the arrival rate during $\Delta t$:
		\begin{align}
		\label{lambda1}
		\lambda_1 = 1.
		\end{align}
		i.e., it is assumed that a single report is generated on average.
		
		In order to model the correlated behavior of the stations, we assume a coupled Markov modulated Poisson process, as proposed in \cite{trafficModels}.
 		The main idea is that a background process $\Theta$ determines the transition probability matrices $\mathbf{P}_n ( t ) $ for all stations $n$, $1 \leq n \leq N $:
		\begin{align}
			\mathbf{P}_n ( t )  = & \begin{pmatrix}
	  			p_{n,00} ( t ) & p_{n,01} ( t ) \\
	  			p_{n,10} ( t )  & p_{n,11}  ( t )
	 		\end{pmatrix} \nonumber \\
			= & \,  (1-\theta_n ( t) ) \cdot \mathbf{P}_0 +  \theta_n ( t ) \cdot \mathbf{P}_1, \label{eq:P}
		\end{align}
		where the samples of the background process $\theta_n(t)$ are in the range $[0,1]$, and where $\mathbf{P}_0$ and $\mathbf{P}_1$ are the transition probabilities matrices in the regular state (state 0) and alarm reporting state (state 1), respectively:
		\[
		\mathbf{P}_0 = 
	 		\begin{pmatrix}
	  			1 & 0  \\
	  			1 & 0  
	 		\end{pmatrix},
	 		\qquad
	 		\mathbf{P}_1 =
	 		\begin{pmatrix}
	  			0 & 1  \\
	  			1 & 0  
	 		\end{pmatrix},
		\]
		In other words, in the regular reporting state a station never reports alarm, while in the alarm reporting state, a station reports a single alarm and goes back to the regular reporting regime.

		Using the fact that the sum of independent Poisson random variables is a Poisson random variable\cite{asimow2010probability}, the aggregated arrival rate $\lambda$ of station $n$ during the interval of duration $T_I = M \Delta t$ is:
		\begin{equation}
				\lambda_n ( T_I) = \sum_{m=0}^{M-1} \lambda_0 \cdot \pi_{n,0}(m \Delta t)+ \lambda_1 \cdot \pi_{n,1}(m \Delta t ),
		\end{equation}
		where $\mathbf{\pi}_n ( t ) = ( \pi_{n,0}  ( t ) \; \pi_{n,1} ( t ) )$ can be obtained from the balance equation $\mathbf{\pi}_n ( t ) = \mathbf{\pi}_n ( t ) \cdot \mathbf{P}_n ( t )$.
		
		Obviously, the samples of background process $\theta_n( t )$ determine the features of the larm reporting on the cell level, see \eqref{eq:P}.
		In \cite{trafficModels}, $\theta_n( t )$  is modeled using Beta distribution with the parameters given in \cite{TR37.8682011}.
		In contrast to this approach, we characterize $\theta_n( t )$ using a physical model of alarm propagation.
		Specifically, we assume that $\theta_n( t )$ can be expressed as: 
		\begin{equation}
				\theta_n (t) = \Psi (d_n) \cdot \delta \left( t_\texttt{a} - \frac{d_n}{v}\right), \; n=1,\dots,N,
			\label{eq:theta}
		\end{equation}
		where $\Psi ( d_n )$ is the spatial correlation factor, $\delta \left( t- \frac{d_n}{v}\right)$ is a delta pulse that describes the propagation of the event that triggered alarm reporting, $t_\texttt{a}$ is the moment of the occurrence of the alarm event, $d_n$ is the stations distance from the epicenter of the alarm event and $v$ is the propagation speed of the alarm event.
		
		\begin{figure}
	  			\centering
	    		\includegraphics[width=\columnwidth]{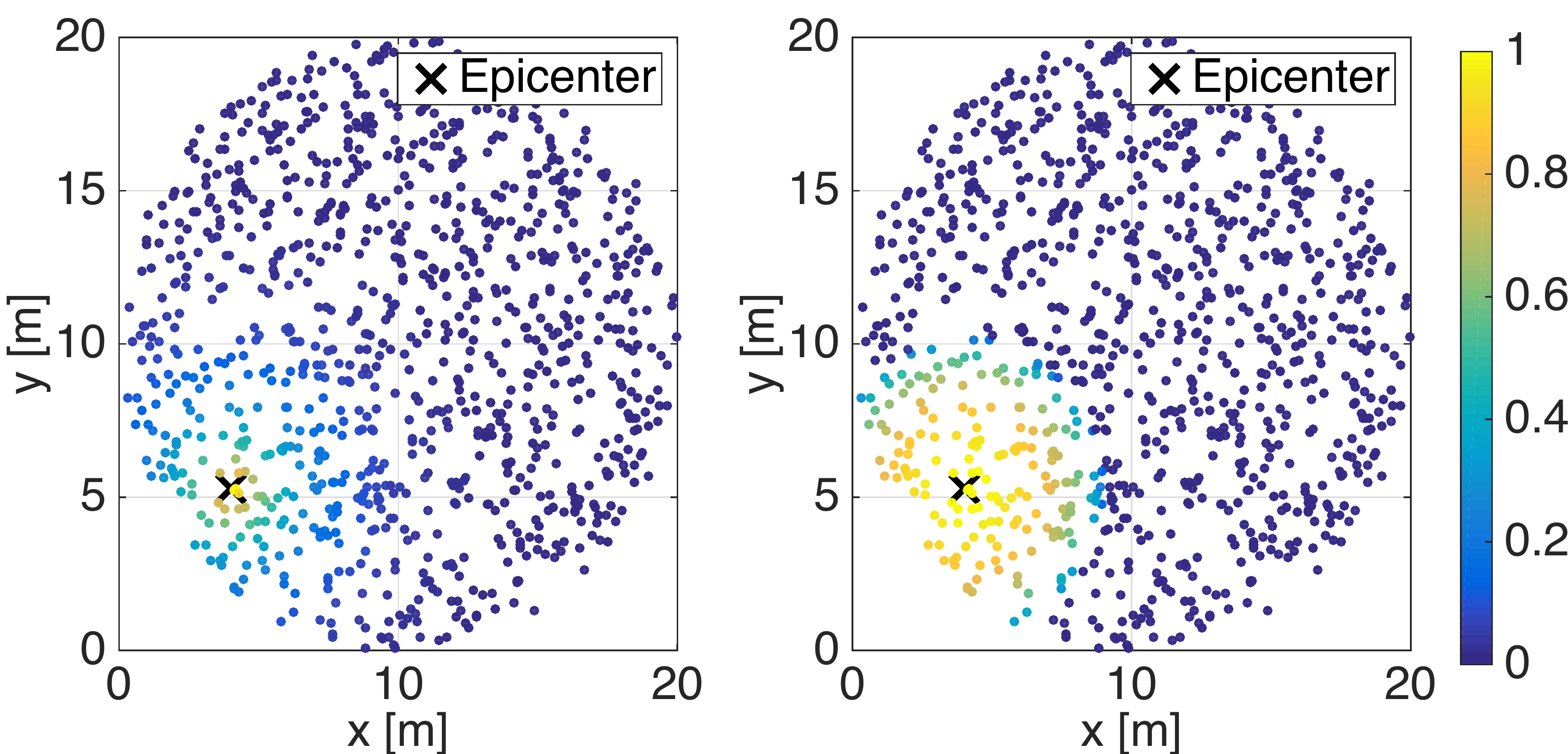}
	    		\caption{Values of spatial correlation factor in a cell with $N=1000$ stations and radius of $r=10 \, \text{m}$. a) Exponentially-decaying spatial-correlation factor $\Psi^\text{Exp}(d)$, $a = 1.2$.  b) Square-root spatial-correlation factor  $\Psi^\text{Sq}(d)$, $d_\text{max}=4 \, \text{m}$.}
	    		\label{fig:correlations}
		\end{figure}
				
			The spatial correlation factor $\Psi ( d )$ expresses probability that a station is affected by an alarm event that took place at distance $d_n$.			
			We consider three different models for $\Psi( d )$.	
			The first model corresponds to the situation where all stations in the cell are affected by the event: 
			\begin{equation}
				\Psi ( d ) = 1.
			\end{equation}
			In the second model, we consider a correlation exponentially decaying with the distance, given as:
			\begin{align}
				 \Psi^\text{Exp}(d)=\left\{
					  \begin{array}{l l}
					  	0, & d < 0,\\
					    e^{-a \cdot d}, &   d \geq 0, 
					  \end{array} \right. \label{expDecay}	
			\end{align}
			where $d$ is the distance of the station to the epicenter of the alarm event and $a$ is the decay constant\footnote{We assume that $a$ is given in $\text{m}^{-1}$.}.
			Finally, as the third model we consider a square root function: 			
			\begin{align}
				 \Psi^\text{Sq}(d)=\left\{
					  \begin{array}{l l}
					  	\sqrt{d_\text{max}^2 - d^2}, &   0 \leq d \leq d_\text{max} ,\\
					        0, & \text{else}, 
					  \end{array} \right. \label{concave}	
			\end{align}
			where $d_\text{max}$ is the maximum distance with respect to the epicenter that the event can reach.
			
			The examples of the exponentially-decaying and the square-root spatial correlation factor are shown in Fig.~\ref{fig:correlations}.
			For the sake of comparison, the decay constant $a$ is chosen such that $\Psi^\text{Sq} (d) < 0.01$ for $d  \geq d_\text{max}$, where $d_\text{max}$ is the parameter that determines the maximum spreading of the alarm for $\Psi^\text{Exp} (d )$.
			Obviously, the values of spatial correlation are higher when the spatial correlation of alarm reporting is modeled by \eqref{concave}.
			
		\begin{figure}[t]
			\centering
	   		\includegraphics[width=0.99\columnwidth]{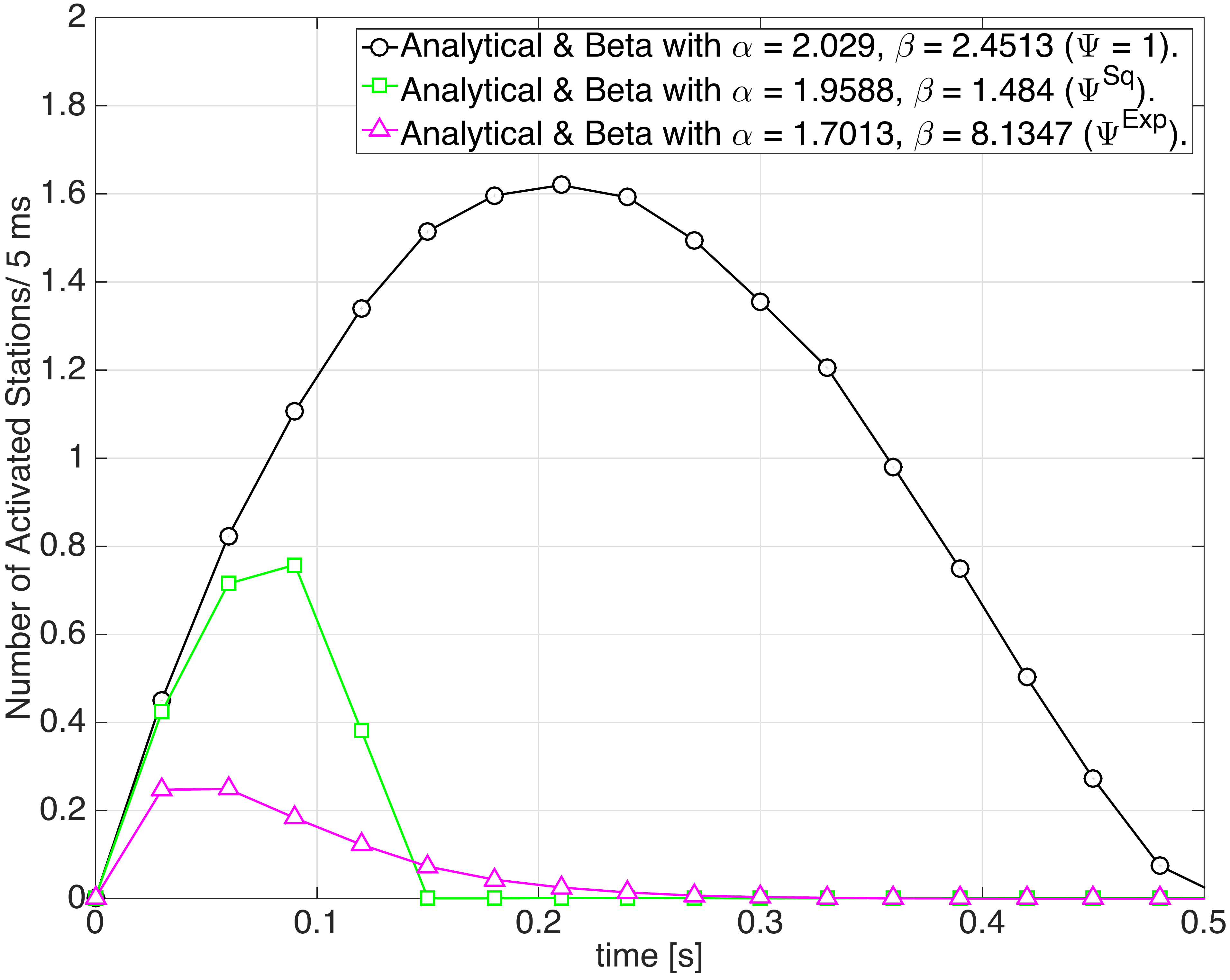}\caption{Evolution of the number of alarm reports over time for three spatial correlation factors: $\Psi ( d ) = 1$, $\Psi^\text{Sq}( d)$ with $d_\text{max} = 500 \, \text{m}$, and $\Psi^\text{Exp}$ with $a = 0.005$. Scenario corresponds to a cell with $r=1000 \, \text{m}$, where the event propagates at $v=4 \, \text{km/s}$.}
	   	\label{fig:beta}
		\end{figure}
		
	\subsection{Illustrative Examples}
	\label{sub:comparison}
		
		We provide a brief illustration of the alarm reporting patterns in a cell for the different correlation factors.
		The example scenario corresponds to a power outage in the grid due to an earthquake, where the primary wave propagates at an average speed $v=4$~km/s \cite{TR37.8682011}, the cell radius is set to $r=1000 \, \text{m}$ and the number of stations is $N=1000$.
		
		Fig.~\ref{fig:beta} shows how the number of alarm reports (i.e., the number of the alarm reporting stations) evolves over intervals of $5 \, \text{ms}$, for the three correlation factors: $\Psi ( d ) = 1$, $\Psi^\text{Sq} ( d ) $ with $d_\text{max}=500 \, \text{m}$ and $\Psi^\text{Exp} ( d ) $ with $a  = 0.005$.
		Obviously, for $\Psi ( d ) = 1$ all stations are triggered and the activation period is the longest.
		The total number of triggered stations for $\Psi^\text{Sq} ( d ) $ is larger than for $\Psi^\text{exp} ( d ) $; however, the activation period is longer for $\Psi^\text{Exp} ( d )$.
		The depicted results were also validated in a simulation setup that implements alarm reporting in a cell with the corresponding number of stations and using the models of the background process \eqref{eq:theta}.

		\begin{figure*}[t]
			\begin{center}
			\includegraphics[width=0.85\textwidth]{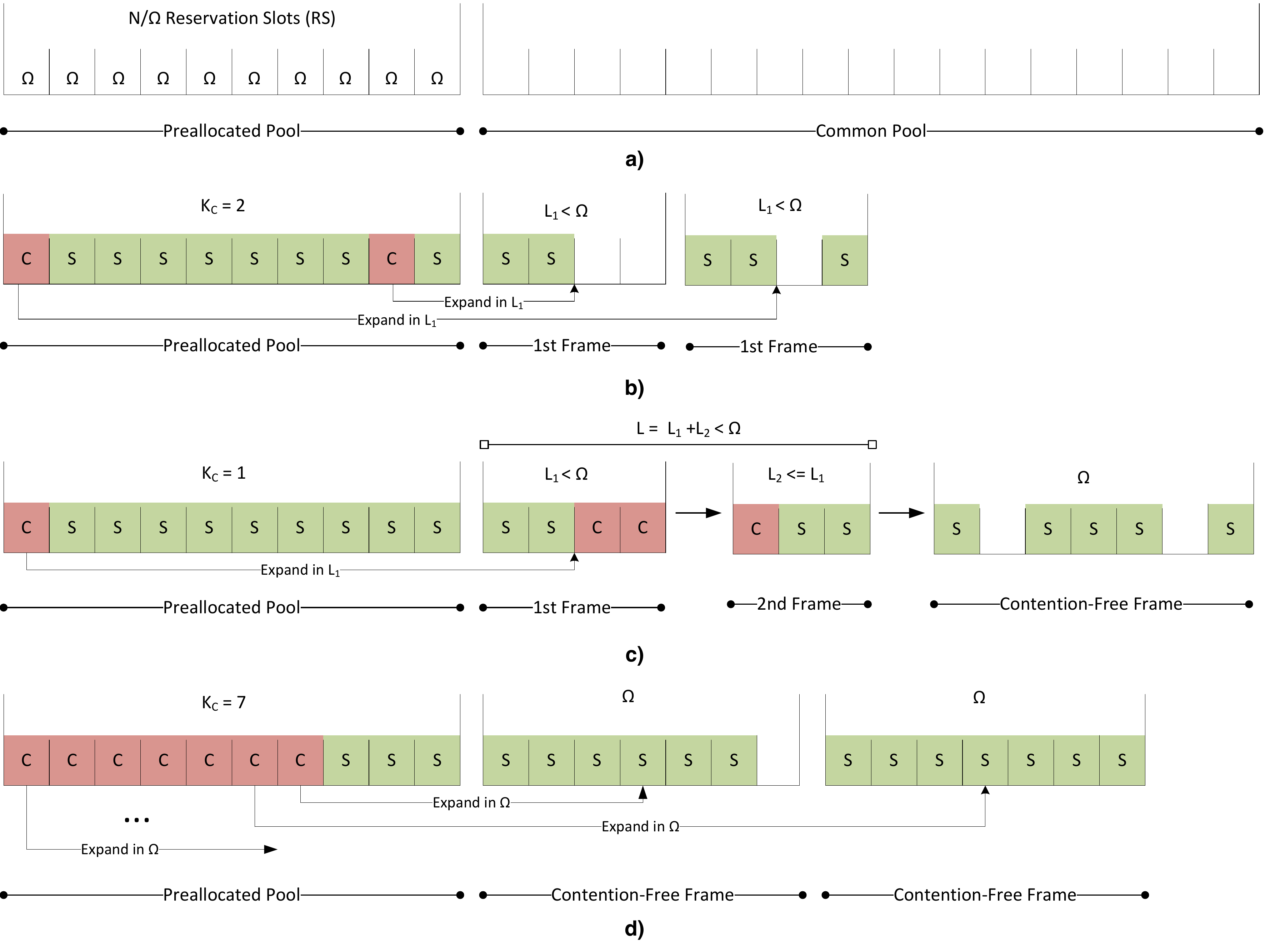}
			\end{center}
			\caption{a) Structure of the pool of resources: the preallocated and the common pool. b) Preallocated pool working under regular regime, where RSs in collision split in $L_1$ extra RS. c)  Preallocated pool working under regular regime, where a collision in the common pool is resolved using frames with $L_1$, $L_2$ and, finally, $\Omega$ RSs. d) Preallocated pool under alarm regime where every collision RS expands into frame with $\Omega$ RSs.}
		\label{fig:reservation-slots}
		\end{figure*}

		Interestingly, it can be shown that all activation patterns showed in Fig.~\ref{fig:beta} can be described by a Beta distribution with the appropriate choice of the shape parameters and activation periods, see \eqref{eq:beta}.
		The corresponding values are indicated in the figure's legend.
		Obviously, these values significantly differ from the ones proposed by 3GPP \cite{TR37.8682011}.
		Specifically, the activation period $T$ is one order magnitude shorter than the value suggested in\cite{TR37.8682011}.
		Also, it could be shown that the number of activated devices over time for the model proposed in \cite{TR37.8682011} is negligible in comparison to the values depicted in Fig.~\ref{fig:beta}.
		We note that the similar conclusions can be drawn for other values of spatial correlation factor.

		The observations above justify the use of Beta distribution to model synchronous reporting by a massive number of stations in a cell.
		However, the parameters of the distribution have to be carefully tuned to the properties of the phenomenon that describes the spreading of the alarm event, and there is no one-size-fits-all solution.
		
		In Section~\ref{sec:results} we demonstrate the performance of the proposed allocation mechanism in the case of an alarm whose spatial correlation factor is modeled by $\Psi^\text{Sq} ( d )$ and note that similar studies based on different spatial correlation factors could be done in an analogous manner.
	
\section{The Proposed Access Method}
	\label{sub:proposedAllocation}

		The main feature of the proposed access method is that it consists of a periodically reoccurring pool of reservation slots (RSs).
		We assume that RS is a RAW slot dimensioned to fit a poll frame that consists of a station's AID and a single bit that indicates that the station has a pending uplink report, which is required by the IEEE 802.11ah protocol.
		The pool is reoccurring with a fixed period $T_\text{R}$.
		The pool duration $T_{\text{pool}}$ is adaptive and determined by the reporting demands of the stations, such that all stations that have pending report and are requesting a dedicated slot are identified.\footnote{We do not consider the potential impact of the noise-induced channel errors in this paper, as our focus is on the operation of the access protocol.}
		The bound on joint size of $T_\text{R}$ and $T_{\text{pool}}$ is posed by the deadline $\tau$ that corresponds to the minimum allowed delay for the identification of the reporting station:	
		\begin{align}
		\label{eq:cond}
		 \tau > T_\text{R} + \max \{ T_{\text{pool}} \},
		 \end{align}
		where in the assumed reporting model $\tau = \tau_\texttt{a}$, see \eqref{eq:deadlines}, and where $T_{\text{pool}}\leq T_\text{R}$, see Fig.~\ref{AID}c).
				
		The pool is divided into two parts: the preallocated and the common pool, as depicted in Fig.~\ref{fig:reservation-slots}a).
		The preallocated pool consists of fixed number of RSs, each one dedicated to $\Omega$ stations.
		A station that experience packet arrivals attempts access in the next dedicated RSs when it occurs, i.e., the access is gated. 
		In a straightforward approach, one could design preallocated pool such that there is a RS dedicated to every station, which corresponds to the case when $\Omega = 1$.
		Although inherently reliable, this solution is very inefficient.
		Typically the stations are only periodically reporting with $T_\text{RI} >> T_\text{R}$, which implies that a station does not have a report to transmit in every pool and, therefore, most of the RSs will be idle (i.e., wasted).\footnote{Specifically, we assume that there is a maximum of a single arrival per station during $T_\text{R}$, and that in regular mode, most of the stations do not experience packet arrivals during $T_\text{R}$. This is justified when \eqref{eq:periodic}, \eqref{eq:deadlines}, \eqref{eq:cond} and the alarm reporting model from Section~\ref{sec:reporting model} are taken into account.}
		The key idea is to preallocate the same RS to a group of $\Omega > 1$ stations, such that the probability of a RS being idle is kept at the tolerable level.
		The group is defined by the stations AIDs, the allocation of the respective RS is announced in the beacon frame, and the total size of the preallocated pool is $ \lceil N / \Omega \rceil$ RSs.
		The access in the preallocated pool is contention based, where in the worst case there could be $\Omega$ stations contending per RS.
		However, as shown in Section~\ref{sec:analysis}, dimensioning of $\Omega$ is also done such that, under regular reporting, there is a low probability that two or more stations from the same group will be active in the same RS and cause a collision.	
				
		The purpose of the common pool is to foster resolution of collisions that occur in the preallocated pool.
		The actual resolution mechanism, as well as the size of the common pool depend on the number of the collisions observed in the preallocated pool $k_C$.
		In particular, $k_C$ is compared to a predefined threshold $\Delta_C$ to decide whether to resolve collisions using a \emph{contention-based }or a \emph{contention-free} approach.	
		If $ k_C < \Delta_C $, the AP assumes that the stations perform only regular reporting.
		For every group of stations that collided in a RS of the preallocated pool, a new frame with $L_1 < \Omega$ RSs is allocated in the common pool, in which the stations from the group contend again by transmitting in a random RSs, see Fig.~\ref{fig:reservation-slots}b). 
		In other words, every collision RS from the preallocated pool expands into frame with $L_1$ slots in the common pool, where the collided stations contend using frame slotted ALOHA approach \cite{OIN1977}.
		If there are collisions in this frame, a second frame with $L_2$ slots is allocated for their resolution, where $L_2 \leq L_1$, in which collided stations contend again using frame slotted ALOHA approach, see Fig.~\ref{fig:reservation-slots}c).\footnote{When the collision resolution is frame slotted Aloha-based, a higher efficiency can achieved using two frames with (optimized) lengths of $L_1$ and $L_2$ slots, respectively, than providing a single frame with  $L= L_1 + L_2$ slots \cite{filterFrame}.}		
		Ultimately, if there are still collisions in the second frame, the AP provides a contention-free frame with $\Omega$ RSs, i.e., a RS dedicated to each, potentially unresolved station from the original group, as shown in Fig.~\ref{fig:reservation-slots}c). 	
		On the other hand, if $k_C \geq \Delta_C$,  the AP immediately opts for a contention-free strategy, assuming that the collisions are due to an ongoing alarm reporting that affects many stations.
		In that case, for every collided RS in the preallocated pool, $\Omega$ RSs are allocated in the common pool, see Fig.~\ref{fig:reservation-slots}d), and each station transmits in its dedicated RS. 
		Finally, once all the active stations have been identified, the AP assigns a data slot for each station, and the operation of the access mechanism continues as in the standard triggered RAW mode.
		
		$\Omega$, $\Delta_C$, $L_1$ and $L_2$ are design parameters.
		In the next section we analyze their optimal dimensioning such that the proposed access method is able to detect if there is on-going alarm reporting in the cell, identify all active stations  and use the resources efficiently.
				
\section{Analysis}
\label{sec:analysis}

	As already elaboreted, the proposed access method tunes its operation to the reporting activity in the cell, i.e., whether there is only the regular or a combination of the regular and alarm reporting in the cell.
	Denote by $H_0$ the hypothesis that there is only regular reporting in the cell, and denote by $H_1$ the alternative hypothesis, i.e., that an alarm event has taken place, affecting a significant number of stations in the cell.
	The threshold $\Delta_c$ is used to determine what is the nature of the reporting activity in the cell, by comparing it to the number of the collisions observed in the periodic pool:
	\begin{align}
    			H_0: k_c < \Delta_c, \\ 
    			H_1: k_c \geq \Delta_c. 
	\end{align}

	Let $C_{ij}$ denote the cost of deciding $H_i$ when $H_j$ is true, $i,j = 0,1$.
	The costs $C_{ij}$, $i,j=0,1$, represent the number of RSs used in the respective modes of operation of the access mechanism, given the actual reporting activity. 
	The total expected cost is:
	\begin{align}
		E[C] &= \sum_{i=0}^1 \sum_{j=0}^1 E[ C_{ij} ] \cdot P(H_i|H_j) \cdot P(H_j),
	\label{eq:expectedCost}
	\end{align}
	where $P(H_j)$, $j=0,1$, denotes the a priori probability of the hypothesis, which can be calculated using, e.g., the knowledge of the history of alarm events:
	\begin{align}
		P(H_1) & = \frac{\text{no. of alarms/day}}{\text{no. of pools/day}}, \\
		P(H_0) & = 1 - P(H_1).
	\end{align}
	
	The probability of correctly deciding that there is only regular reporting activity in the cell is:
	\begin{align}
		& P(H_0 | H_0)  =   P ( k_C  < \Delta_C | H_0 ) \nonumber \\
		\label{eq:ak}
		=   \sum_{k=0}^{\Delta_c -1} &  \binom{\lceil N / \Omega\rceil}{k} \cdot P_{C_{H0}}^k \cdot (1-P_{C_{H0}})^{\lceil N / \Omega\rceil - k},
	\end{align}
	where $\lceil N / \Omega \rceil$ is the number of RSs in the preallocated pool and $P_{C_{H0}}$ denotes the probability of a collision in a RS in the regular reporting regime.
	We assume in \eqref{eq:ak} that the contention outcomes over RSs independent identically distributed (i.i.d.) random variables; this assumption was verified by the evaluation that simulates an IEEE 802.11ah cell and the proposed reporting setup, see Section~\ref{sec:results}.
	The probability of false alarm detection $P(H_1|H_0)$ is simply:
	\begin{equation}
		P(H_1 | H_0) = 1 - P(H_0 | H_0).
	\end{equation}
	The probability of detecting the alarm is:
	\begin{align}
		P(H_1 | H_1)   & = P (k_C \geq \Delta_C | H_1 ) \nonumber \\
		=   \sum_{k = \Delta_c}^{ \lceil N / \Omega\rceil } & \binom{\lceil N / \Omega\rceil}{k} \cdot P_{C_{H1}}^k \cdot (1-P_{C_{H1}})^{\lceil N / \Omega\rceil - k},
	\end{align}
	where $P_{C_{H1}}$ denotes the probability of collision in a RS in the alarm regime.
	Again, we assume that the contention outcomes over RSs are (i.i.d.) random variables, which was verified by the evaluation.
	Finally, the probability of alarm miss $P(H_0 | H_1)$ is:
	\begin{equation}
		P(H_0|H_1) = 1 - P(H_1|H_1).
	\end{equation}

	\subsection{Derivation of Collision Probabilities}
	\label{sub:probabilityCollision}

	The probabilities of collision $P_{C_{H0}}$ and $P_{C_{H1}}$ are the probabilities that there are at least two stations that are active in the same RS:
		\begin{align}
			\label{eq:PCH0}
			P_{C_{H0}}  = 1 - (1-p_{A,0})^{\Omega} - \Omega \cdot  p_{A,0} \cdot (1-p_{A,0})^{\Omega-1}, \\
			\label{eq:PCH1}
			P_{C_{H1}} = 1 - (1-p_{A,1})^{\Omega} - \Omega \cdot  p_{A,1} \cdot (1-p_{A,1})^{\Omega-1}.
		\end{align}
		where $p_{A,0}$ and $p_{A,1}$ indicate the probabilities of a station reporting when in state 0/state 1, respectively:
		\begin{align}
			p_{A,0} = 1 - e^{-\lambda_0 \cdot T_\text{R}}, \\
			P_{A,1} = 1 - e^{- \lambda (T_\text{R}) },
		\end{align}
		see Section~\ref{sec:reporting model}.
	
	\subsection{Derivation of Costs}
	\label{sub:costs}
		
		Consider first the cost $C_{00}$, i.e., the number of RSs used in the contention-based operation of the access method when there is only regular reporting in the cell. 
		$C_{00}$ is computed as:
		\begin{equation}
			C_{00} = \lceil N / \Omega \rceil + \sum_{i=1} ^ {k_{C_{00}}} S_{i},
		\end{equation}
		where the first term on the right is the size of the preallocated poll in RSs, while the second term corresponds to the size of the common pool.
		Specifically, $k_{C_{00}}$ is the number of RSs in collision in the preallocated pool, while $S_i$ denotes number of RSs needed to resolve $i$-th collision from the preallocated pool.
		$k_{C_{00}}$ and  $S_i$, $i = 1, \dots, k_{C_{00}} $, are independent random variables, while the latter are also identically distributed. 
		Using Wald's equation \cite{wald}, the expected value $E[C_{00}]$ can be computed as:
		\begin{align}
		E [ C_{00} ] = \lceil N / \Omega \rceil + E [ k_{C_{00}} ] \, E [ S ].
		\end{align}
		We proceed by computation of $E [ S ]$, while derivation of $ E [ k_{C_{00}} ] $ is presented in Section~\ref{sec:ks}.
		
		Recall from Section~\ref{sub:proposedAllocation} that every RS in collision from the preallocated pool expands first into a frame with $L_1$ RSs, then in a second frame with $L_2$ RSs if the collision is not resolved in the first frame, and finally, in a contention-free frame with $\Omega$ RSs if the collision is not resolved in the second frame.
		It is easy to verify that:
		\begin{align}
		E [ S ] = & L_1  + L_2 ( 1 - R_1 ) + \Omega ( 1 - ( R_1 + R_2 ) ), 
		\end{align}
	where $R_{1}$ is the probability of resolving the collision in the first frame and $R_{2}$ is the probability of resolving the collision in the second frame. 

		We proceed by deriving $R_{1}$ and $R_{2}$.		
		Denote by $R ( h | m , L )$ the conditional probability that $h$ users are resolved when $m$ users contend in the frame of length $L$, assuming frame slotted ALOHA-based contention.
		It could be shown that \cite{vogt2002efficient}:
		\begin{align}
		R ( h | m , L ) = \frac{ { L_1 \choose h } }{  L ^ m }  G ( L - h , m - h ) \prod_{i=0}^{h - 1} ( m - i ) ,
		\end{align}
		where:
		\begin{equation}
			G(u,v) = u^v + \sum_{t=1}^{v} (-1)^t \prod_j^{t-1}[(v-j)(u-j)](u-t)^{v-t}\frac{1}{t!}.
		\end{equation}	
		$R_1$ can be computed as:
		\begin{align}
		\label{eq:R1}
		R_1 = \sum_{m = 0}^{\Omega} R ( m | m , L_1 ) P_{A} ( m ),
		\end{align}
		where $P_{A} ( m )$ denotes the probability that $m$ stations are contending in the first frame.
		$P_{A} ( m )$ can be computed as:
		\begin{align}
		P_{A} ( m ) = \left\{
					  \begin{array}{l l}
					  	\frac{\binom{\Omega}{k}  p_{A,0}^k  (1-p_{A,0})^{\Omega-k}}{ F( 2 , \Omega, p_{A,0} )}, & 2 \leq m \leq \Omega, \\
					  	0, & \text{ else},
					  \end{array} \right. \label{truncated}	
		\end{align}
		where:
		\begin{align}
		\label{eq:F}
		F ( l, v, u ) = \sum_{t  = l}^{v}  \binom{v}{t}  u^t  (1- u )^{v -  t}. 
		\end{align}
		In other words, \eqref{truncated} expresses the fact that there has to be at least two colliding stations if there was a collision.
		Analogously as in \eqref{eq:R1}, $R_{2}$ can be computed as:
		\begin{equation}
			R_{2} =  \sum_{m=0}^{\Omega} \sum_{h = 2}^{m} R ( h | h, L_2 ) R ( m - h | m, L_1 ) P_{A} ( m ).
		\end{equation}		
		
		$E [ C_{10} ] $, i.e., the expected cost when operating in the contention-free mode when there are no alarms is simply:
		\begin{equation}
			E [ C_{10} ] = \lceil N / \Omega \rceil + E [ k_{C_{10}} ] \, \Omega,
		\end{equation}
		where $ E [ k_{C_{10}} ] $ is the expected number of collision RSs in the preallocated pool, and each of these expands into frame with $\Omega$ RSs.
		
		Further, $ E [ C_{01} ] $, i.e., the cost of operating in the contention-based mode when there are alarms, is:
		\begin{equation}
			E [ C_{01} ] =  \lceil N / \Omega \rceil + E [ k_{C_{01}} ]  \left( L_1 + L_2 + \Omega \right),
		\end{equation}
		where we assumed that the probability of resolving collision RSs in the preallocated pool using the first and the second frame in the common pool is negligible, such that every RS in collision is finally resolved using a contention-free frame of $\Omega$ RSs.
		
		Finally, $E [ C_{11} ] $, i.e., the expected cost of operating in contention-free mode when there is an alarm is:
		\begin{align}
			E [ C_{11} ] =  \lceil N / \Omega \rceil + E [ k_{C_{11}} ] \, \Omega. 
		\end{align}
	
		\subsection{Expected Number of Collision RSs}
		\label{sec:ks}
		
		We proceed by deriving the expected number of collision RSs, required for the computation of the expected costs.
		It can be shown that:
		\begin{align}
			E [ k_{C_{00}} ] & = \frac{ \sum_{k=0}^{\Delta_C -1} k \cdot \binom{\lceil\frac{N}{\Omega}\rceil}{k} \cdot P_{C_{H0}}^k \cdot (1-P_{C_{H0}})^{\lceil\frac{N}{\Omega}\rceil-k}}{ F (0, \Delta_C - 1, P_{C_{H0}} ) },\\
			E [ k_{C_{10}} ] & = \frac{ \sum_{k=\Delta_C -1}^{\lceil\frac{N}{\Omega}\rceil} k \cdot \binom{\lceil\frac{N}{\Omega}\rceil}{k} \cdot P_{C_{H0}}^k \cdot (1-P_{C_{H0}})^{\lceil\frac{N}{\Omega}\rceil-k}}{F (\Delta_C, \lceil\frac{N}{\Omega}\rceil , P_{C_{H0}} )}, \\
			E [ k_{C_{01}} ] &  = \frac{ \sum_{k=0}^{\Delta_C -1} k \cdot \binom{\lceil\frac{N}{\Omega}\rceil}{k} \cdot P_{C_{H1}}^k \cdot (1-P_{C_{H1}})^{\lceil\frac{N}{\Omega}\rceil-k}}{F (0, \Delta_C - 1, P_{C_{H1}} )}, \\
			E [ k_{C_{11}} ] & = \frac{ \sum_{k=\Delta_C -1}^{\lceil\frac{N}{\Omega}\rceil} k \cdot \binom{\lceil\frac{N}{\Omega}\rceil}{k} \cdot P_{C_{H1}}^k \cdot (1-P_{C_{H1}})^{\lceil\frac{N}{\Omega}\rceil-k}}{ F (\Delta_C, \lceil\frac{N}{\Omega}\rceil , P_{C_{H1}} )},
		\end{align}
		where $\Delta_C$ is the detection threshold, $P_{C_{H0}}$, $P_{C_{H1}}$ and $F ( l, u, v )$ are given in \eqref{eq:PCH0}, \eqref{eq:PCH1} and \eqref{eq:F}, respectively, and where we assumed that the contention outcomes over RSs are i.i.d. random variables.

\section{Evaluation}
\label{sec:results}

		\begin{figure}[t]
		\centering
	   	\includegraphics[width=0.95\columnwidth]{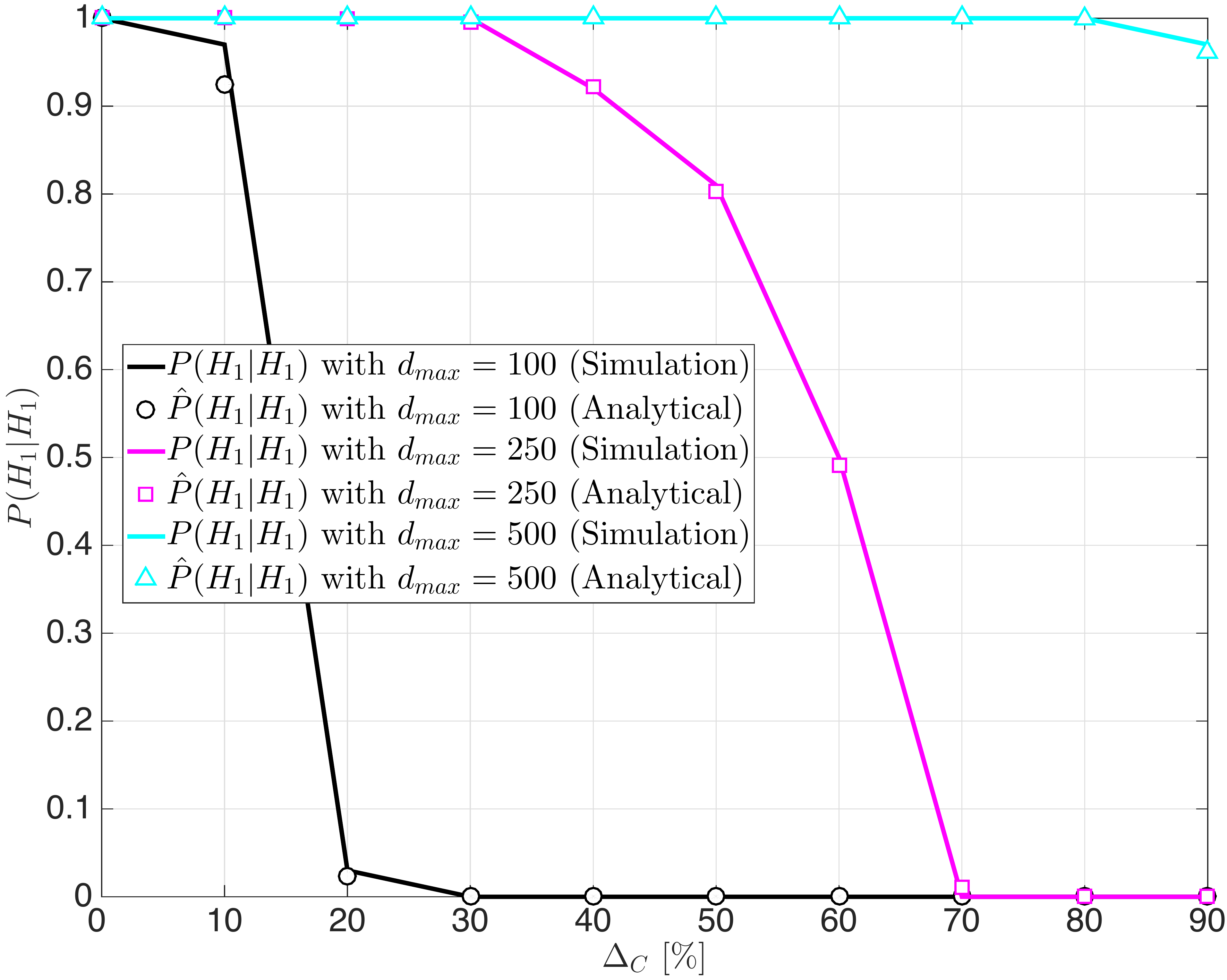}\caption{$P ( H_1 | H_1 )$ in a 1000-m radius cell with $N=8000$ stations and $\Omega = 40$. The alarm threshold $\Delta_C$ is given as the percentage of RSs in collision in the preallocated pool.}
	   	\label{fig:probDetection}
	\end{figure}

	In this section we evaluate the proposed solution in terms of the expected costs $E[C]$ and the probability of alarm detection $P ( H_1 | H_1 )$.
	The investigated scenario corresponds to a heavily loaded case with 8000 smart meters in a 1000-m radius cell, close to the maximum capacity of a IEEE 802.11ah cell (i.e., 8100 stations).
	The traffic model of smart meters comprises periodic reporting, on-demand reporting and alarm reporting.
	For the periodic reporting we use the reporting interval $T_\text{RI}=5 \, \text{min}$, thus the corresponding arrival rate is $\lambda_\texttt{p} = 300^{-1}$~report/s, which is a typical value \cite{Madueno2017}.
	For  the on-demand reporting, RI is set to 15~min, i.e.,  $\lambda_\texttt{d} = 1500^{-1}$~report/s, which is a rather demanding configuration in comparison with typical values \cite{openSmartGrid}.
	For the alarm reporting we assume a scenario where the alarm propagates at an average speed of $v=4$~km/s \cite{TR37.8682011}, which corresponds to a power outage or power fluctuation in the grid due to, e.g., an earthquake, while the correlation between stations with respect to the distance is given by the function $\Psi^{Sq} ( d ) $, see (\ref{concave}).
	According to a relevant description of smart grid traffic messages from \cite{openSmartGrid}, we assume that a maximum tolerable delay for alarm is $\tau = 5 \, \text{s}$.
	We adopt a rather conservative approach and illustrate the performance when the pool reoccurring period is set to $T_\text{R} = 2.5 \, \text{s}$, which allows for the maximum pool duration of $\max{T_\text{pool}}=2.5$~s, see~\eqref{eq:cond}.
	RS duration is set to $200 \, \mu\text{s}$, which, at the lowest modulation scheme, is enough to fit the transmission of the 13~bits AID and an uplink data flag bit, and leaves room for the transmission of the additional 23 bits that may be devoted to other information,  e.g., a CRC trailer, etc.	
	Finally, we note that 	all the analytical results presented in this section were verified using simulations of a IEEE 802.11ah cell with the same communication setup.

	\begin{figure*}[t]
  		\centering
    	\includegraphics[width=0.7\textwidth]{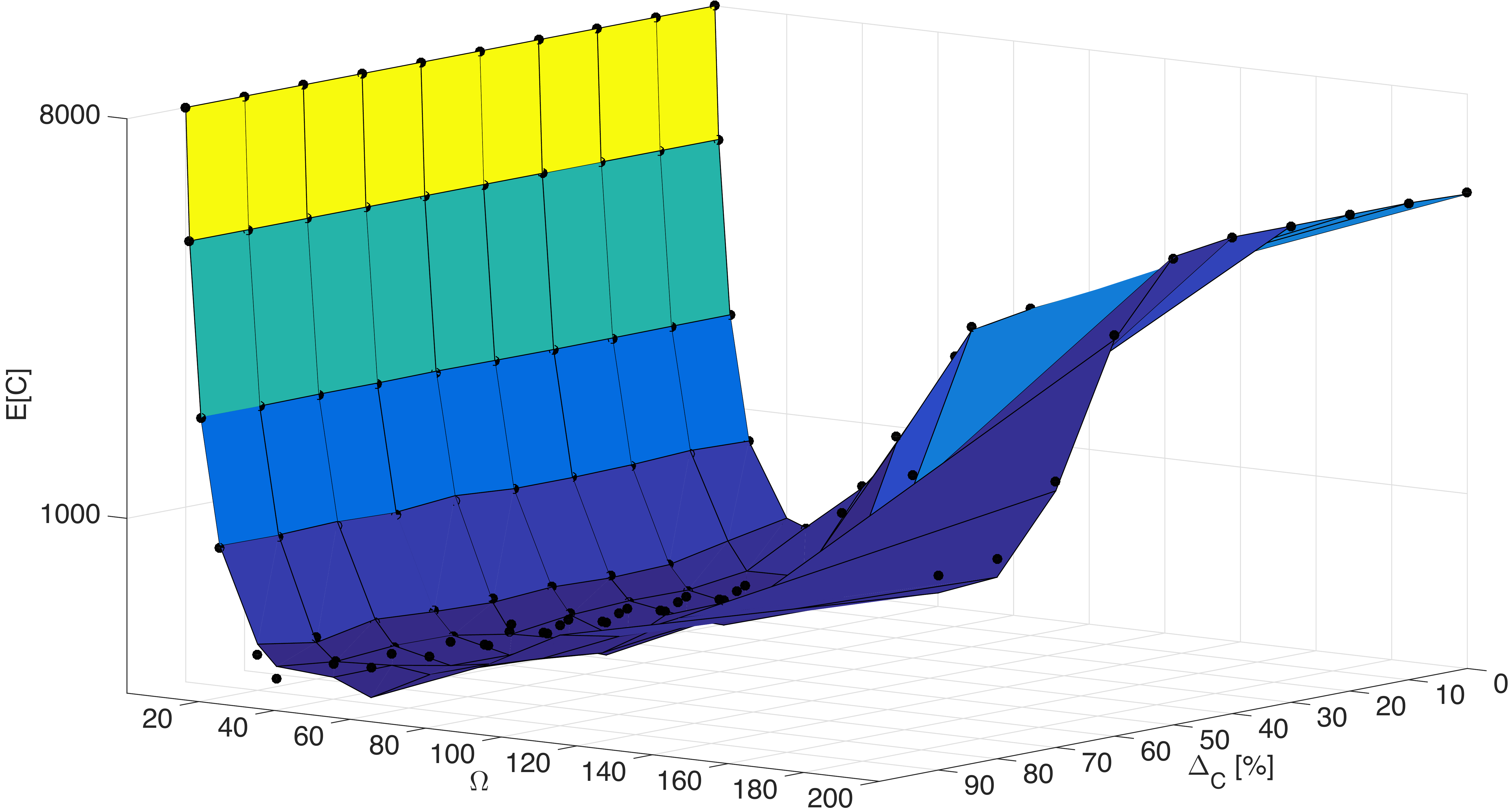}
    	\caption{Expected cost $E[C]$  with respect to the alarm threshold $\Delta_C$ as percentage of total RSs in collision and the slot degree $\Omega$, where $L_1 = 0.6 \, \Omega$, $L_2=0.4 \, \Omega$ and the alarm probability $P(H_1) = 5 \cdot 10^{-3}$; the dots indicate analytical results and the solid lines simulation results.}
    	\label{fig:E_C}
	\end{figure*}
	
	Fig.~\ref{fig:probDetection} illustrates $P(H_1|H_1)$ as function of the alarm detection threshold $\Delta_C$, for varying maximum propagation distance from the epicenter $d_\text{max}$.
	Obviously, $\Delta_C$ has to be selected according to $d_{max}$, as $d_{max}$ affects the number of alarm reporting stations.
	For example, if we aim to detect an alarm event where $d_\text{max} = 250$~m, $\Delta_C$ should be set below $30$\% of RSs in collision, and when $d_\text{max} = 500$~m, the alarm is always detected if $\Delta_C$ is set below $80$\% of RSs in collision.	

	Next, we investigate the efficiency of the proposed mechanism as function of the alarm threshold $\Delta_C$ and RS degree $\Omega$.
	We assume that the frames sizes used for collision resolution in the common pool are $L_1 = 0.6 \, \Omega$ and $L_2 = 0.4 \, \Omega$; these values were obtained through an exhaustive search optimization that bests fit the arrival rates of the stations and statistics of the number of collision RSs in the preallocated pool in the regular reporting regime.
	Further, we assume the probability of alarm occurrence is $P ( H1 ) = 5 \cdot 10^{-3}$.
	Fig.~\ref{fig:E_C} illustrates the expected cost $E [ C ]$ per pool as function of $\Omega$ and  $\Delta_C$.
	Obviously, for low values of $\Omega$  the expected cost $E[C]$ is high; this is due to the fact the for low $\Omega$ the number of RSs in the preallocated pool is overdimensioned with respect to the reporting activity in the cell and many of these RSs are idle.
	The worst case is for $\Omega=1$, when every station has a reserved RS per pool.
	Increasing $\Omega$ sharply reduces the cost, until the breakpoint at roughly around $\Omega=100$ after which too many collisions start to occur, requiring an increased number of resources in the common pool to resolve them. 
	The alarm threshold $\Delta_C$ also plays an important role in the efficiency of the system.
	If $\Delta_C$ is set too low, probability of false alarm detection increases, incurring the waste of resources in the common pool, as every collision RS in the preallocated is automatically expanded in a frame of $\Omega$ RSs.
	On the other hand, with a high $\Delta_C$ the probability of alarm miss increases, which also leads to the waste of resources in the common pool.
	Specifically, in this case there is a high chance that a collision RS from the preallocated pool has to be expanded in frames with $L_1$, $L_2$ and $\Omega$ RSs.
	A closer inspection of the results depicted in Fig.~\ref{fig:E_C} reveals that the optimal configuration is $\Omega=40$ and $\Delta_C =  50$\% of total RSs in collision, where $E[C]\approx 400$ RSs.
	
\begin{figure}[t]
  		\centering
    	\includegraphics[width=0.95\columnwidth]{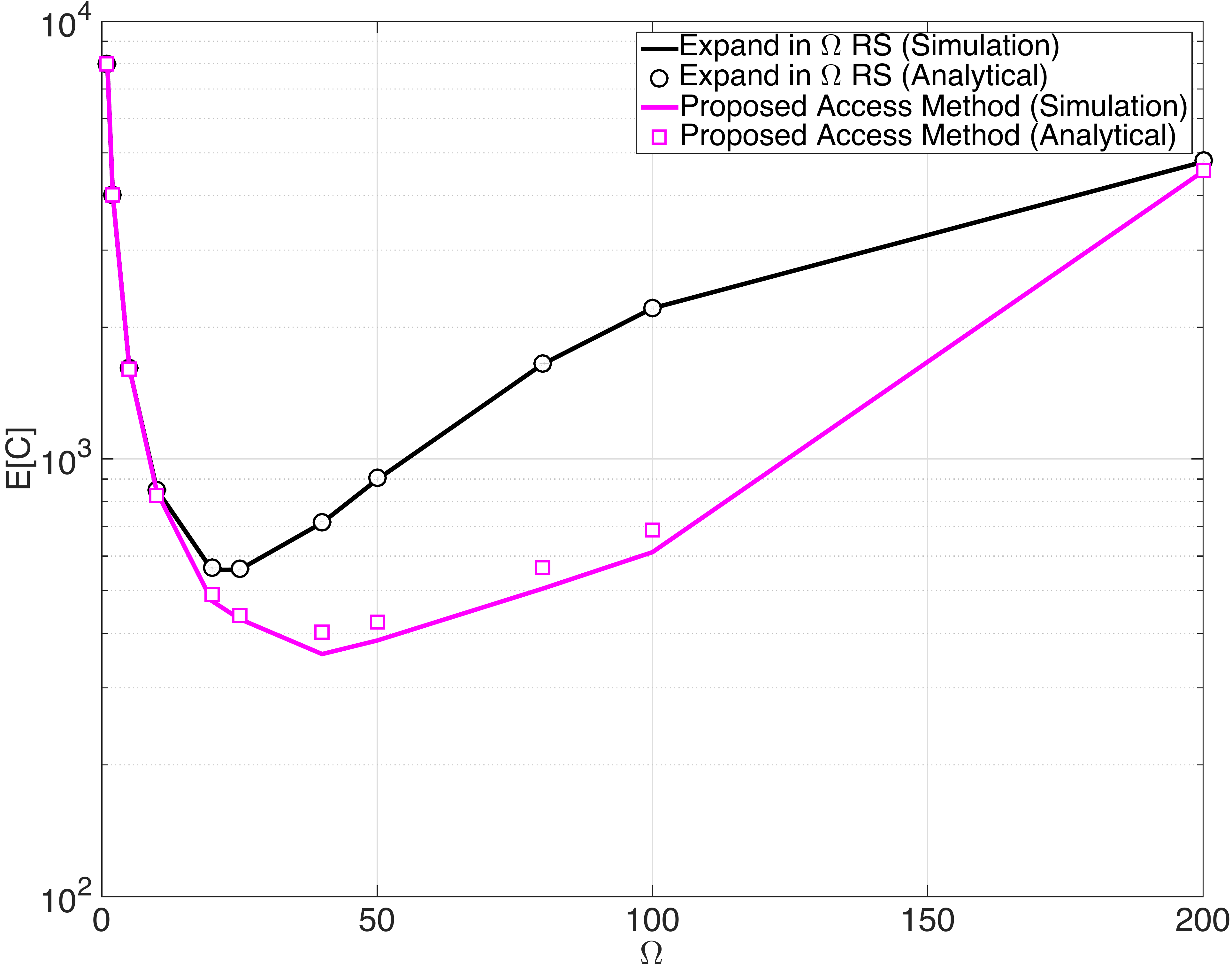}
    	\caption{Expected cost $E[C]$ vs. $\Omega$, for $\Delta_C = 50 \, \%$ of RSs in collision and $P(H_1) = 5 \cdot 10^{-3}$.}
    	\label{fig:CostComparison}
	\end{figure}

	A more detailed view of the configuration corresponding to $\Delta_C =  50$\% of total RSs in collision is shown in Fig.~\ref{fig:CostComparison}.
	For the sake of comparison we have also included the expected cost of the naive approach in which each collision RS in preallocated pool is always expanded into $\Omega$ additional RSs in the common pool (i.e., the access mechanism in the common pool operates only in the contention-free mode).
	Obviously, the proposed scheme is less sensitive in the variations in $\Omega$ and exhibits a superior performance, justifying the use of the contention-based collision resolution in the common pool.
	Specifically, when the corresponding minimum expected costs are compared, the naive approach requires in average roughly 2 times more RSs than the proposed scheme.	
	
	We conclude by commenting on the reliability/de-lay/efficiency tradeoff that the proposed access method offers.
	Taking into account the assumed regular reporting interval of $5 \, \text{min}$ and the period of the pool reoccurring of $T_\text{R} = 2.5 \, \text{s}$ in the considered scenario, there are 120 pools during the regular reporting interval.
	When the optimal configuration parameters $\Omega=40$ and $\Delta_C =  50$\% are taken into account, the expected cost per pool is $E[C] \approx 400$ RSs, and the pool duration is just $T_\text{pool} = 80 \, \text{ms}$.
	Also, on average there are 6 RS are spend per station during the interval of $5 \, \text{min}$, i.e., in all 120 pools, while at the same time, all the reporting stations are always resolved within the reporting deadline of $\tau = 5 \, \text{s}$.
	For the simple polling mechanism that ensures that the same deadline is met, which is for $\Omega = 1$, the cost per pool is 8000 RSs and the pool duration is fixed to $T_\text{pool} = 1.6 \, \text{s}$, and there are 120 RSs per station during the regular reporting interval.
	The proposed access method could be further optimized, i.e., $T_\text{RI}$ could be prolonged and the number of pools reduced, by taking into account the statistics of the pool duration conditioned $T_\text{RI}$ and dimensioning $T_\text{RI}$ such that \eqref{eq:cond} is satisfied.
	However, we leave such a study for further work.
	Finally, we note that, if the duration of RS should be increased in order to accommodate longer messages (which could include, e.g., short message identification or other information elements, besides AID and the uplink flag), this reduction in number of required RSs becomes even more important, as it directly influences the pool duration and hence pool reoccurring interval to achieve the reporting deadlines.
	
\section{Discussion and Conclusions}
\label{sec:conclusion}
	
	In this paper we proposed an access mechanism tailored for M2M reporting applications in IEEE 802.11ah networking scenario.
	Its main features are tuning of the operation to the reporting activity in the call by proactively dimensioning the number of used resources, i.e., slots, while resolving all reporting stations within the given reporting deadline.
	The proposed solution can be seamlessly incorporated in the triggered RAW operation of IEEE 802.11ah access protocol.
	
	As a side result, we investigated the modeling of  alarm reporting in a IEEE 802.11ah cell, using spatio-temporal models of a alarm propagation and provided justifications for the use of the Beta distribution to model inter-arrival times of alarm reporting.
	To the best of our knowledge, such a study does not exist in the previous literature.
		
	Finally, we note that the we assumed a static, i.e., fixed placement of the M2M terminals in the cell.
	However, this assumption is not restrictive and the presented results could be applied to the cases when the terminals are mobile.
	In particular,  the terminal mobility within the cell does not affect the regular reporting model, and, for the alarm reporting model, the spreading of alarms is typically several orders of magnitude faster than the potential velocity of M2M terminals.

\bibliographystyle{IEEEtran}

\begin{thebibliography}{10}
\providecommand{\url}[1]{#1}
\csname url@samestyle\endcsname
\providecommand{\newblock}{\relax}
\providecommand{\bibinfo}[2]{#2}
\providecommand{\BIBentrySTDinterwordspacing}{\spaceskip=0pt\relax}
\providecommand{\BIBentryALTinterwordstretchfactor}{4}
\providecommand{\BIBentryALTinterwordspacing}{\spaceskip=\fontdimen2\font plus
\BIBentryALTinterwordstretchfactor\fontdimen3\font minus
  \fontdimen4\font\relax}
\providecommand{\BIBforeignlanguage}[2]{{%
\expandafter\ifx\csname l@#1\endcsname\relax
\typeout{** WARNING: IEEEtran.bst: No hyphenation pattern has been}%
\typeout{** loaded for the language `#1'. Using the pattern for}%
\typeout{** the default language instead.}%
\else
\language=\csname l@#1\endcsname
\fi
#2}}
\providecommand{\BIBdecl}{\relax}
\BIBdecl

\bibitem{qualcommProjects}
\BIBentryALTinterwordspacing
Qualcomm, \emph{White Paper 802.11ah}, 2015 (accessed in July 15, 2015).
  [Online]. Available:
  \url{https://www.qualcomm.com/invention/research/projects/wi-fi-evolution/80211ah}
\BIBentrySTDinterwordspacing

\bibitem{capacityAnalysis}
T.~Adame, A.~Bel, B.~Bellalta, J.~Barcelo, J.~Gonzalez, and M.~Oliver,
  ``{Capacity analysis of IEEE 802.11 ah WLANs for M2M communications},'' in
  \emph{Multiple Access Communications}, ser. Lecture Notes in Computer
  Science.\hskip 1em plus 0.5em minus 0.4em\relax Springer, 2013, vol. 8310,
  pp. 139--155.

\bibitem{groupStrategy}
L.~Zheng, L.~Cai, J.~Pan, and M.~Ni, ``Performance analysis of grouping
  strategy for dense ieee 802.11 networks,'' in \emph{Proc. IEEE GLOBECOM
  2013}, Atlanta, GA, USA, Dec. 2013.

\bibitem{estimationRAW}
C.~W. Park, D.~Hwang, and T.-J. Lee, ``Enhancement of ieee 802.11ah mac for m2m
  communications,'' \emph{IEEE. Commun. Letters}, vol.~18, no.~7, pp.
  1151--1154, Jul. 2014.

\bibitem{Madueno2017}
G.~{Corrales Madue\~no}, C.~Stefanovic, and P.~Popovski, ``{Reengineering
  GSM/GPRS Towards a Dedicated Network for Massive Smart Metering},'' in
  \emph{Proc. IEEE SmartGridComm 2014}, Venice, Italy, Nov. 2014.

\bibitem{lastGasp}
{IEEE 802.16p}, ``{IEEE 802.16p Machine to Machine (M2M) Evaluation Methodology
  Document (EMD)},'' {IEEE 802.16 Broadband Wireless Access Working Group
  (802.16p)}, EMD {11/0005}, 2011.

\bibitem{WCL}
G.~Madue\~no, C.~Stefanovic, and P.~Popovski, ``{Reliable Reporting for Massive
  M2M Communications With Periodic Resource Pooling},'' \emph{IEEE Wireless
  Commun. Letters}, vol.~3, no.~4, pp. 429--432, Aug. 2014.

\bibitem{TR37.8682011}
3GPP, ``{Study on RAN Improvements for Machine-type Communications},'' {3rd
  Generation Partnership Project (3GPP)}, TR {37.868 V11.0}, Aug. 2010.

\bibitem{uplinkAccess}
{IEEE Task Group ah (TGah)}, ``Uplink channel access general procedure,'' 2012.

\bibitem{R1975}
L.~G. Roberts, ``Aloha packet system with and without slots and capture,''
  \emph{SIGCOMM Comput. Commun. Rev.}, vol.~5, no.~2, pp. 28--42, Apr. 1975.

\bibitem{openSmartGrid}
{Open Smart Grid SG-Network task force}, ``Smart grid networks system
  requirements specification, release version 5,'' 2010.

\bibitem{betaFunction}
G.~Andrews, R.~Askey, and R.~Roy, \emph{Special Functions}.\hskip 1em plus
  0.5em minus 0.4em\relax Cambridge University Press, 2000.

\bibitem{trafficModels}
M.~Laner, P.~Svoboda, N.~Nikaein, and M.~Rupp, ``Traffic models for machine
  type communications,'' in \emph{Proc. ISWCS 2013}, Ilmenau, Germany, Aug.
  2013, pp. 1--5.

\bibitem{asimow2010probability}
L.~Asimow and M.~Maxwell, \emph{Probability and Statistics with Applications: A
  Problem Solving Text}, ser. ACTEX academic series.\hskip 1em plus 0.5em minus
  0.4em\relax ACTEX Publications, 2010.

\bibitem{OIN1977}
H.~Okada, Y.~Igarashi, and Y.~Nakanishi, ``Analysis and application of framed
  {ALOHA} channel in satellite packet switching networks - {FADRA} method,''
  \emph{Electronics and Communications in Japan}, vol.~60, pp. 60--72, Aug.
  1977.

\bibitem{filterFrame}
G.~{Corrales Madue\~no}, N.~K. Pratas, C.~Stefanovic, and P.~Popovski,
  ``{Massive M2M Access with Reliability Guarantees in LTE Systems},'' in
  \emph{Proc. IEEE ICC 2015}, London, UK, Jun. 2015.

\bibitem{wald}
S.~M. Ross, \emph{Introduction to probability and statistics for engineers and
  scientists}.\hskip 1em plus 0.5em minus 0.4em\relax Academic Press, 2004.

\bibitem{vogt2002efficient}
H.~Vogt, ``{Efficient Object Identification with Passive RFID Tags},'' in
  \emph{Pervasive Computing}, ser. Lecture Notes on Computer Science.\hskip 1em
  plus 0.5em minus 0.4em\relax Springer, 2002, vol. 2414, pp. 98--113.

\end{thebibliography}

\end{document}